\documentclass[default,iicol]{sn-jnl}

\usepackage{makecell}
\usepackage{epstopdf}
\usepackage{float}
\usepackage[utf8]{inputenc}

\begin{document}

\title[]{Exploring the Possibility of Interacting Quintessence Model as an Alternative to the $\Lambda$CDM Model}


\author*[]{\fnm{Nandan} \sur{Roy}}\email{nandan.roy@mahidol.ac.th}



\affil[]{\orgdiv{Centre for Theoretical Physics and Natural Philosophy}, \orgname{Mahidol University}, \orgaddress{\street{Nakhonsawan Campus}, \city{Phayuha Khiri, Nakhonsawan}, \postcode{60130}, \country{Thailand}}}




\abstract{This study examines interacting quintessence dark energy models and their observational constraints for a general parameterization of the quintessence potential, which encompasses a broad range of popular potentials. Four different forms of interactions are considered. The analysis is done by expressing the system as a set of autonomous equations for each interaction. The Bayesian Model Comparison has been used to compare these models with the standard Lambda Cold Dark Matter ($\Lambda$CDM) model. Our analysis shows positive and moderate evidence for the interacting models over the $\Lambda$CDM model. }


\maketitle

\section{Introduction}

The accelerated expansion of the universe has been confirmed by various cosmological observations \cite{SupernovaSearchTeam:1998fmf, SupernovaCosmologyProject:1998vns, Meszaros:2002np, Planck:2014loa, ahn2012ninth} but the reason behind it remains a mystery. Cosmological constant \cite{padmanabhan2006dark} is considered as the simplest and most successful candidate for dark energy, but it still faces major theoretical challenges like the cosmological constant problem and the coincidence problem.

Recent high-precision cosmological data has shown a statistically significant discrepancy in the estimation of the current value of the Hubble parameter ($H_0$) between early-time and late-time observations, which poses another challenge to the cosmological constant. Early universe measurements like CMB Planck collaboration \cite{Planck2020} (including BAO \cite{BAO2017, BAO2011}, BBN \cite{BBN2021}) and DES \cite{DES2018,DES:2018rjw,krause2017dark} collaboration estimate $H_0 \sim (67.0 - 68.5)$ km/s/Mpc, while late-time distance ladder measurements like SH0ES \cite{Sh0ES2019} and H0LiCOW \cite{Wong:2019kwg} collaborations using time-delay cosmography method report $H_0 = (74.03 \pm 1.42)$ km/s/Mpc.  Over the years this discrepancy has increased of the order of $\simeq 5.3 \sigma$\cite{Riess_2022}, indicating the possibility of new physics  beyond $\Lambda$CDM in the dark energy sector.

Dynamical dark energy models, such as quintessence, k-essence, phantom dark energy, etc., have been proposed as alternatives to the cosmological constant \cite{amendola2010dark,Bamba:2012cp}. These models involve a scalar field with a potential energy that drives the accelerated expansion of the universe, and the equation of state of the dark energy evolves with time \cite{copeland2006dynamics,Peebles2003,Armendariz2001,roy2022quintessence,Banerjee:2020xcn,Lee:2022cyh,Krishnan:2020vaf}. The possibility of interactions between dark matter and dark energy in dynamical dark energy models is not ruled out from both theoretical and observational perspectives. Interactions between the dark sectors have been shown to alleviate the cosmic coincidence problem\cite{Cai:2004dk,mangano2003coupled,Sadjadi:2006qp,Wang:2016lxa,Jesus:2020tby}, and in recent years, interacting models have gained attention for their potential to resolve the $H_0$ and $\sigma_8$ tensions \cite{Salvatelli2014,Costa2017,DiValentino2019,Kumar2020,DiValentino:2019ffd,DiValentino:2019jae,Yang:2018euj,Wang:2018duq}.

In cosmology, the interaction between dark matter and dark energy is considered by introducing some unknown interaction terms into the continuity equation. It has been proposed that the dark matter and dark energy components are not conserved separately but instead conserved jointly. The form of the interaction is arbitrary and is generally chosen based on its phenomenological performance. The consideration of the interaction between the dark sectors should affect the expansion history and overall evolution of the universe\cite{Wang:2016lxa}. Numerous studies have extensively explored the cosmological ramifications arising from the interactions between dark matter and dark energy \cite{amendola2000coupled, farrar2004interacting, mangano2003coupled, tamanini2015phenomenological, chimento2010linear, pan2015analytic, pettorino2005extended, pettorino2008coupled}. These investigations encompass a broad range of perspectives, including both theoretical considerations and observational analyses. 

Dynamical systems analysis has been extensively used to examine the qualitative behavior of various cosmological models, including interacting dark energy models. Generally, one can convert the Einstein field equations, along with the interaction term, into a set of autonomous equations and employ dynamical systems analysis techniques to investigate the stability of these models. Previous studies have already explored models with different types of interactions, encompassing both general relativity and modified gravity models \cite{Khyllep:2021wjd,Caldera-Cabral:2008yyo,Amendola:1999er,Boehmer:2008av,Zonunmawia:2017ofc,Hussain:2022dhp}. For a comprehensive understanding of the dynamical systems analysis of interacting dark energy models, we recommend referring to the following review: \cite{Bahamonde:2017ize}.

In this study, we analyzed the performance of interacting quintessence dark energy models using state-of-the-art cosmological data at the background level. We considered a very general setup of the quintessence potential by using the parametrization of the potential from \cite{Roy:2018nce}, which includes a large class of potentials. Four distinct interaction terms were considered, and the Einstein field equations for the interacting quintessence field were reformulated into a set of autonomous equations through appropriate variable transformations. The models were then implemented in the Boltzmann code CLASS and evaluated against recent cosmological observations using the MCMC code Montepython. We employed the concept of Bayes factor and Jeffreys scale to compare these interacting models with each other and also with the $\Lambda CDM$  model.

 \par The structure of the present study is outlined as follows: In Section \ref{mathematicalbg}, we provide an overview of the mathematical formulation and the dynamics of the scalar field. Section \ref{interaction} focuses on the mathematical setup for each type of interaction term considered. The initial conditions and the implementation of the model in the CLASS code, as well as the constraints obtained from recent cosmological observations, are described in Section \ref{numerical}. Finally, our results and findings are summarized in Section \ref{conclusion}.

\section{Mathematical Background} \label{mathematicalbg}
 
Let us consider a spatially flat Friedmann-Robertson-Walker (FRW) universe that is composed of radiation, dark matter, and dark energy, with the latter two components interacting with each other. We consider quintessence scalar field as our chosen dark energy component and we further assume that the components of the universe are barotropic in nature and obey the relation $p_j = w_j \rho_j$, where $w_r = 1/3$ for radiation and $w_m = 0$ for dark matter. For the above-mentioned universe, the Einstein field equations are written as
\begin{subequations}
\label{eq:field}
  \begin{eqnarray}
    H^2 &=& \frac{\kappa^2}{3} \left( \sum_j \rho_j +
      \rho_\phi \right) \, , \label{eq:field1} \\
    \dot{H} &=& - \frac{\kappa^2}{2} \left[ \sum_j (\rho_j +
      p_j ) + (\rho_\phi + p_\phi) \right] \,,
      \label{eq:field2} 
  \end{eqnarray}
\end{subequations}
where, $\kappa^2 = 8 \pi G$ and $a$ is the scale factor of the Universe, while $H \equiv \dot{a}/a$ denotes the Hubble parameter, and the dot represents the derivative with respect to cosmic time. The continuity equations for each component, including radiation, matter, and the scalar field, can be expressed as follows:

\begin{subequations}
\begin{eqnarray} \label{eq.conty}
\dot{\rho_r} + 3 H \rho_r (1 + w_r) &=& 0, \label{con_rad} \\ 
\dot{\rho_m} + 3 H \rho_m (1 + w_m) &=& - Q ,  \label{con_mat}\\
\dot{\rho_\phi} + 3 H (\rho_\phi + p_\phi) &=& + Q. \label{con_phi}
\end{eqnarray}
\end{subequations}
In this context, $Q$ denotes the coupling between the quintessence field and the matter sector, and we have chosen a convention such that if $Q$ is positive, the energy transfer occurs from dark matter to dark energy, whereas if $Q$ is negative, the energy transfer occurs from dark energy to dark matter. The densities of radiation, matter, and the scalar field are represented by $\rho_r$, $\rho_m$, and $\rho_\phi$, respectively, while their equation of state (EoS) is denoted as $w_r$, $w_m$, and $w_\phi$. The wave equation for the scalar field can be expressed as follows:
\begin{equation} \label{eq:wave}
\ddot{\phi} + 3 H \dot{\phi} + \frac{dV(\phi)}{d\phi} = \frac{Q}{\dot \phi},
\end{equation}

where the potential of the scalar field is $V(\phi)$.

To write down the evolution equations of the quintessence field as an set of autonomous equation, we introduce the following set of dimensionless variables,

\begin{subequations}
  \label{eq:4}
  \begin{eqnarray}
    x &\equiv& \frac{\kappa \dot{\phi}}{\sqrt{6} H} = \Omega_{\phi} ^{1/2} \sin(\theta/2),\label{eq:4a} \\ 
   y & \equiv&  \frac{\kappa V^{1/2}}{\sqrt{3} H} = \Omega_{\phi}^{1/2}
\cos(\theta/2 \, , \label{eq:3a}
    \\
    y_1 &\equiv& - 2\sqrt{2} \frac{\partial_{\phi} V^{1/2}}{H} , \\
    y_2 &\equiv& - 4\sqrt{3} \frac{\partial^2_\phi
                 V^{1/2}}{\kappa H} \, . \label{eq:3b}             
  \end{eqnarray}
\end{subequations} 

This particular transformation was first used in \cite{Urena-Lopez:2015gur} and later it is used in \cite{Roy:2018nce}. Using these sets of new variables the system of equations that governs the dynamics of the scalar field reduces to the following set of autonomous equations,

\begin{subequations}
\label{eq:polar}
 \begin{eqnarray}
    \theta^{\prime} &=& - 3 \sin \theta + y_1 + q \ \Omega_{\phi} ^{1/2} \cos(\theta/2)\, , \label{eq:10a} \\
    y_1^{\prime} &=& \frac{3}{2} \left( 1  + w_{tot} \right) y_1
                     + \Omega_{\phi} ^{1/2} \sin(\theta /2) y_2  \,
                     , \label{eq:10b} \\ 
    \Omega_{\phi} ^{\prime} &=& 3 (w_{tot} - w_{\phi})
                                \Omega_{\phi} + q \ \Omega_{\phi} ^{1/2} \sin(\theta/2) \, , \label{eq:10c}       
  \end{eqnarray}
\end{subequations}

where $q = \frac{\kappa Q}{\sqrt[]{6} H^2 \dot{\phi}}$ represents the interaction in the new system and a `prime' is the differentiation with respect to the e-foldings $N = \ln(a)$. From now on we consider the unit $\kappa^2 =1$. The total EoS of the system is given as follows,
\begin{equation}
 w_{tot} \equiv \frac{p_{tot}}{\rho_{tot}} = \sum _{i}w_{i}\Omega _{i} = \frac{1}{3} \Omega_r + \Omega_\phi w_\phi.
\end{equation}

Here $\Omega_\phi$ is the scalar field energy density parameter and the scalar field EoS is given by $w_\phi = -\cos\theta$. One can notice the system of equations in Eq.(\ref{eq:polar}) is not closed until unless one consider a particular form of the $y_2$. In this work we will be considering the following form of the $y_2$;

\begin{equation} 
y_2 = y \left( \alpha_0 + \alpha_1 y_1/y + \alpha_2 y^2_1/y^2 \right) \, . \label{eq:GP1}
\end{equation}

This form of $y_2$~\cite{Roy:2018nce,Urena-Lopez:2020npg} includes a large number of popular scalar field potentials and particularly important as one can study different classes of scalar field solutions without considering any particular form of the potential. The $\alpha$ parameters in the above expression of $y_2$ are called the active parameters which affect the dynamics of the scalar field.

\section{The interaction} \label{interaction}

Although current observations allow for the possibility of an interaction between dark matter and dark energy, the precise form of this interaction remains unknown. In this study, we have explored the following four different forms of interactions (i) $Q=\beta \rho_m \dot{\phi}$, (ii) $Q=\beta \rho_\phi \dot{\phi}$, (iii) $Q=\beta \rho_m \sqrt{\rho_\phi} \dot{\phi}/H$, and (iv) $Q=\beta \dot{\phi} H^2 w_\phi$, here $\beta$ is the coupling parameter. The choices of these specific forms of interaction are phenomenological and also intended to simplify the mathematics and facilitate the closure of the autonomous systems in Eq.(\ref{eq:polar}).

\subsection{Interaction I ($Q=\beta \rho_m \dot{\phi}$)}

 This interaction was first used in \cite{Wetterich:1994bg} and later in \cite{amendola2000coupled}. In this  particular form of interaction the autonomous system of the scalar field dynamics will be reduced to the following;

\begin{subequations}
\label{eq:int1}
  \begin{eqnarray}
    \theta^{\prime} &=& - 3 \sin \theta + y_1 \nonumber \\
    &+&\sqrt{3/2} \beta \Omega_{\phi}^{1/2} (1 - \Omega_\phi) \cos(\theta/2)\, ,  \\
    y_1^{\prime} &=& \frac{3}{2} \left( 1  + w_{tot} \right) y_1
                     + \Omega_{\phi} ^{1/2} \sin(\theta /2) y_2  \,
                     ,  \\ 
    \Omega_{\phi} ^{\prime} &=& 3 (w_{tot} - w_{\phi})
                                \Omega_{\phi} \nonumber \\
                                &+& \sqrt{3/2} \beta \Omega_{\phi}^{1/2} (1 - \Omega_\phi) \sin(\theta/2) \, .       
  \end{eqnarray}
\end{subequations}

The continuity equation of the matter sector in terms of the variables given in Eq.(\ref{eq:4}) can be written as 

\begin{equation} \label{eq:int1_rhom}
    \dot{\rho_m} + 3 H \rho_m = - \sqrt{6}\beta H \rho_m \sqrt{\Omega_\phi} \sin{\theta/2}
\end{equation}

\subsection{Interaction II ($Q=\beta \rho_\phi \dot{\phi}$)}
Though not exact a similar form has been considered in \cite{Kumar:2019wfs} (see the references there in also). For this particular choice of interaction the autonomous system reduces to the following;

\begin{subequations}
\label{eq:int2}
  \begin{eqnarray}
    \theta^{\prime} &=& - 3 \sin \theta + y_1 \nonumber \\
    &+&\sqrt{3/2} \beta \Omega_{\phi}^{3/2}  \cos(\theta/2)\, ,  \\
    y_1^{\prime} &=& \frac{3}{2} \left( 1  + w_{tot} \right) y_1
                     + \Omega_{\phi} ^{1/2} \sin(\theta /2) y_2  \,
                     ,  \\ 
    \Omega_{\phi} ^{\prime} &=& 3 (w_{tot} - w_{\phi})
                                \Omega_{\phi} \nonumber \\
                                &+& \sqrt{3/2} \beta \Omega_{\phi}^{3/2} \sin(\theta/2) \, .        
  \end{eqnarray}
\end{subequations}

Similar to the previous case the continuity equation of the matter sector reduces to the following; 

\begin{equation} \label{eq:int2_rhom}
    \dot{\rho_m} + 3 H \rho_m = -3 \sqrt{6} \beta  H^3 \sin(\theta/2) \Omega _{\phi }^{3/2}
\end{equation}

\subsection{Interaction III ($Q=\beta \rho_m \sqrt{\rho_\phi} \dot{\phi}/H$)}
 The autonomous system reduces to the following for this particular choice of interaction;
\begin{subequations}
\label{eq:int3}
  \begin{eqnarray}
    \theta^{\prime} &=& - 3 \sin \theta + y_1 \nonumber \\
    &+&\frac{3}{\sqrt{2}} \beta \Omega_{\phi} (1-\Omega_\phi)  \cos(\theta/2)\, , \label{eq:11a} \\
    y_1^{\prime} &=& \frac{3}{2} \left( 1  + w_{tot} \right) y_1
                     + \Omega_{\phi} ^{1/2} \sin(\theta /2) y_2  \,
                     , \label{eq:11b} \\ 
    \Omega_{\phi} ^{\prime} &=& 3 (w_{tot} - w_{\phi})
                                \Omega_{\phi} \nonumber \\
                                &+& 3/\sqrt{2} \beta \Omega_{\phi} (1-\Omega_\phi) \sin(\theta/2) \, . \label{eq:11c}       
  \end{eqnarray}
\end{subequations}

The continuity equation for the matter sector can be written as 

\begin{equation} \label{eq:int2_rhom}
    \dot{\rho_m} + 3 H \rho_m = - 3\sqrt{2}\beta H \rho_m \Omega_\phi \sin{\theta/2}.
\end{equation}

\subsection{Interaction IV ($Q=\beta  \dot{\phi} H^2 w_\phi$)}
In this case, the autonomous system is reduced to the following;
\begin{subequations}
\label{eq:int4}
  \begin{eqnarray}
    \theta^{\prime} &=& - 3 \sin \theta + y_1 \nonumber \\
    &-&\frac{1}{\sqrt{6}} \beta \sqrt{\Omega_{\phi}} \cos(\theta)  \cos(\theta/2)\, , \label{eq:11a} \\
    y_1^{\prime} &=& \frac{3}{2} \left( 1  + w_{tot} \right) y_1
                     + \Omega_{\phi} ^{1/2} \sin(\theta /2) y_2  \,
                     , \label{eq:11b} \\ 
    \Omega_{\phi} ^{\prime} &=& 3 (w_{tot} - w_{\phi})
                                \Omega_{\phi} \nonumber \\
                                &-& \frac{1}{\sqrt{6}} \beta \sqrt{\Omega_{\phi}} \cos(\theta)  \sin(\theta/2) \, . \label{eq:11c}       
  \end{eqnarray}
\end{subequations}

and the matter continuity equation takes the following form,

\begin{equation} \label{eq:int4_rhom}
    \dot{\rho_m} + 3 H \rho_m =  \sqrt{6}\beta H^3 \sqrt{ \Omega_\phi} \cos(\theta)\sin{\theta/2}.
\end{equation}

\begin{figure}[h!]
    \centering
    \includegraphics[width=\columnwidth]{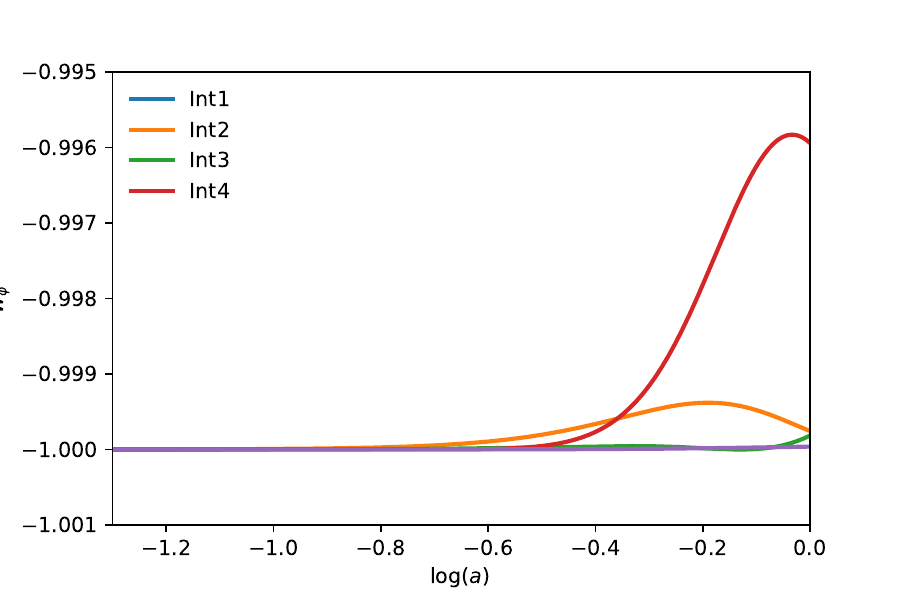}
    \caption{Plot of the evolution of the scalar field EoS($w_{\phi}$) for different interaction forms with $\alpha_i=0$ and other parameters set to their best fit value (see Table.\ref{Tab:constraint}).}
    
    \label{fig:eos}
\end{figure}

In Figure \ref{fig:eos}, we present the evolution of the equation of state (EOS) of the scalar field ($w_\phi$) using the best-fit value of the $\beta$ parameter obtained from the MCMC analysis (see Section \ref{sec:bestfit}) for each interacting model. We consider a general choice of $\alpha_0=\alpha_1=\alpha_1=0$. This particular choice of the $\alpha$ parameters corresponds to the potential taking the form $V(\phi) = (A + B \phi)^2$ (refer to Table II in \cite{Roy:2018nce}). However, it's important to note that different values of the $\alpha$ parameters can be selected, leading to distinct forms of the potential.

From the plots in Figure \ref{fig:eos}, it can be observed that for all the interaction models, the evolution of the EOS is indistinguishable from that of a cosmological constant in the early times. However, at late times, the EOS deviates from the value $w_{\phi} = -1$.

\begin{figure}[h!]
    \centering
    \includegraphics[width=\columnwidth]{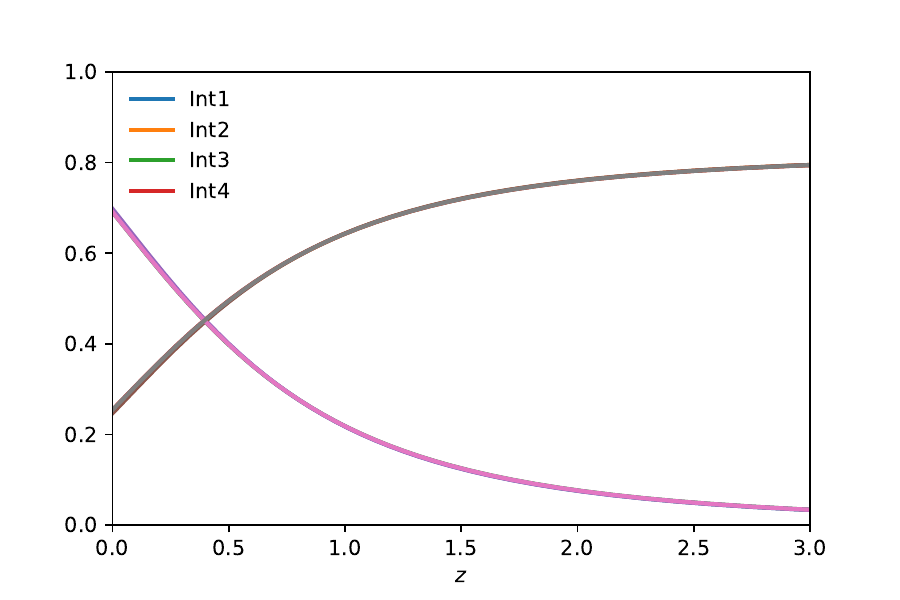}
    \caption{Plot of the $\Omega_m$ and $\Omega_\phi$ for the interacting models for  $\alpha_i =0 $  and other parameters set to their best fit value (see Table.\ref{Tab:constraint}) .}
    \label{fig:density}
\end{figure}

In Figure \ref{fig:density}, we illustrate the evolution of the density parameters $\Omega_m$ and $\Omega_\phi$ using the same choice of $\alpha$ parameters as in Figure \ref{fig:eos}. It is noteworthy that, for all the interacting models, the evolution of $\Omega_m$ and $\Omega_\phi$ cannot be distinguished from each other.

\section{Numerical Simulations} \label{numerical}

\subsection{Initial Condition}
To study the dynamics of the model numerically, we have implemented all the corresponding sets of autonomous equations for the four different interactions, along with the modified continuity equations, in the Boltzmann code CLASS \cite{Lesgourgues:2011rg,Blas:2011rf,Lesgourgues:2011rh}. To obtain reliable numerical solutions, it is important to provide a good guess for the initial conditions to the CLASS code. Following the prescription given in \cite{Roy:2018eug}, we have estimated the initial conditions based on two different assumptions. The first assumption is that the dark energy equation of state is approximately $w_\phi \simeq -1$, leading to $\theta < 1$. The second assumption is that the contribution of the dark energy density during both matter and radiation domination is negligible, i.e., $\Omega_\phi \ll 1$. By considering approximate solutions during radiation and matter domination, and equating them at the radiation-matter equality epoch, we can obtain;

\begin{subequations}
\label{eq:initial_condition}
\begin{eqnarray}
  \theta_i &\simeq& \frac{9}{10} a^2_i \,
  \frac{\Omega^{1/2}_{m0}}{\Omega^{1/2}_{r0}} \theta_0
  \, , \label{eq:7a} \\
  \Omega_{\phi i} &\simeq& a_i^4 \, \frac{\Omega_{m0}}{\Omega_{r0}} \, \Omega_{\phi 0} \, , \label{eq:7b}
\end{eqnarray}
\end{subequations}

The initial value of the variable $y_1$ is related to the angular variable as $y_{1 i} = 5 \theta_i$. The initial condition for $\rho_m$ is taken as $\rho_{mi} = \rho_{m0}(\frac{a_i}{a_0})^{-3}$. In the CLASS code, $a_i$ is typically considered to be $\simeq 10^{-15}$.                                                                                                                                                                                                                                                                         

\subsection{Observational Data}

We utilized the MCMC parameter estimation code Montepython \cite{Brinckmann:2018cvx} to constrain the cosmological parameters. The data sets used for this purpose are the following;

\subsubsection{SN-Ia data}
The type Ia supernovae are commonly acknowledged as standard candles. Their relatively uniform absolute luminosity makes them extremely useful for measuring cosmological distances\cite{reiss1998supernova,SupernovaSearchTeam:1998fmf}. Here we have used the Pantheon compilation sample of SN-Ia data, which was compiled in \cite{Scolnic:2017caz} and includes 1048 data points for SN-Ia. The complete numerical data for the Pantheon SN-Ia catalog is publicly accessible\footnote{http://dx.doi.org/10.17909/T95Q4X} \footnote{https://archive.stsci.edu/prepds/ps1cosmo/index.html}.

The redshift range for Pantheon samples is $0<z<2.3$.  The distance modulus $\mu(z)$ of any type Ia Supernova located at a distance of redshift $z$ is given as $\mu(z) = m - M$, where $m$ represents its apparent magnitude and $M$ is the absolute magnitude. Theoretically, the distance modulus can be written as

$$
\mu_{t h}(z)=5 \log _{10} \frac{d_L(z)}{\left(H_0 / c\right) M p c}+25 ,
$$
where  $H_0$ is the current Hubble rate, $c$ is the speed of light and $d_L(z)$ is the luminosity distance. The luminosity distance $d_L(z)$ in a spatially flat FRW universe is defined as
$$
d_L(z)=(1+z) H_0 \int_0^z \frac{d z^{\prime}}{H\left(z^{\prime}\right)}.
$$

The chi-square of the SN-Ia measurements,
$$
\chi_{S N}^2=\Delta \mu^T \cdot \mathbf{C}_{S N}^{-1} \cdot \Delta \mu .
$$
$\mathbf{C}_{S N}$ is a covariance matrix, and $\Delta \mu=$ $\mu_{o b s}-\mu_{t h}$,  $\mu_{o b s}$ corresponds to the measured distance modulus of a particular SNIa.  Now the distance modulus can be estimated from  the observation of light curves using the empirical formula,

$$
\mu_{obs} = m_B - M + \alpha X_1 - \beta C + \Delta_M + \Delta_B,
$$

Here, $m_B$ is the observed peak magnitude of the SNIa in the rest frame of the B band, and $M$ is the absolute B-band magnitude of a fiducial SNIa. The parameters $\alpha$ and $\beta$ are coefficients that relate the luminosity of the SNIa to its time stretching ($X_1$) and color ($C$), respectively. Additionally, $\Delta_M$ is a distance correction based on the host-galaxy mass of the SNIa, and $\Delta_B$ is a correction based on predicted biases from simulation.

Moreover, the total covariance matrix $\mathbf{C}_{S N}$ is defined as the sum of the statistical matrix $D_{stat}$ and the systematic matrix $C_{sys}$, as shown by the equation
$$
\mathbf{C}_{S N}=D_{\text {stat }}+C_{\text {sys }}
$$
The statistical matrix $D_{stat}$ is a diagonal matrix that contains the distance error of each supernova type Ia (SNIa) along its main diagonal. The distance error is composed of several sources of uncertainty, such as the photometric error, the mass step correction, the peculiar velocity and redshift measurement, the gravitational lensing, the intrinsic scatter, and the distance bias correction. These sources of uncertainty are represented by the terms $\sigma_N^2$, $\sigma_{\text {Mass }}^2$, $\sigma_{\mu-z}^2$, $\sigma_{\text {lens }}^2$, $\sigma_{\text {int }}^2$, and $\sigma_{\text {Bias }}^2$, respectively, in the equation
$$
\sigma^2=\sigma_N^2+\sigma_{\text {Mass }}^2+\sigma_{\mu-z}^2+\sigma_{\text {lens }}^2+\sigma_{\text {int }}^2+\sigma_{\text {Bias }}^2,
$$
The systematic matrix $C_{sys}$ is a non-diagonal matrix that captures the correlation between different SNIa due to systematic effects. The details of how to construct this matrix can be found in Ref.[\cite{Scolnic:2017caz}].

\subsubsection{Baryon Acoustic Oscillation}

Baryon acoustic oscillations are recurring and periodic fluctuations in the density of visible baryonic matter. These oscillations are considered as "standard rulers" for the measurement of distances in cosmology. We employ data points from the following to constraint the cosmological parameters; 

\begin{enumerate}
    \item BOSS DR12 \cite{Alam_2017} at $z = 0.38, 0.51, 0.61$.
    \item eBOSS DR14 (Lya) Combined \cite{deSainteAgathe:2019voe,Cuceu_2019} at $z=2.34$.
    \item WiggleZ Dark Energy Survey \cite{Kazin_2014} at $z= 0.44, 0.6$ and $0.73$.
\end{enumerate}








The Baryon Acoustic Oscillation (BAO) data provides valuable information about the angular diameter distance $d_{\mathrm{A}}(z)$ and the Hubble parameter $H(z)$. To obtain these values, we use the ratio $d_z$, which is defined as follows:

$$
d_z \equiv \frac{r_{\mathrm{s}}\left(z_{\mathrm{d}}\right)}{D_V(z)}
$$

where $D_V(z)$ represents the volume-averaged distance and is given by the expression:

$$
D_V(z)=\left[(1+z)^2 d_{\mathrm{A}}^2(z) \frac{c z}{H(z)}\right]^{1 / 3}
$$

The quantity $r_{\mathrm{s}}\left(z_{\mathrm{d}}\right)$ represents the comoving sound horizon at the drag epoch, and is defined as:

$$
r_{\mathrm{s}}\left(z_{\mathrm{d}}\right)=\frac{1}{H_0} \int_{z_{\mathrm{d}}}^{\infty} \frac{c_{\mathrm{s}}(z)}{H(z) / H_0} \mathrm{~d} z
$$

Here, $c_{\mathrm{s}}(z)$ denotes the sound speed, and $z_{\mathrm{d}}$ is the redshift at the drag epoch and for  the $\Lambda CDM$  model, above equation can be approximated as\cite{1998ApJ496605E,EUCLID:2020syl}:

$$
r_{\mathrm{s}}\left(z_{\mathrm{d}}\right) \simeq \frac{44.5 \log \left(\frac{9.83}{\Omega_{\mathrm{m}, 0} h^2}\right)}{\sqrt{1+10\left(\Omega_{\mathrm{b}, 0} h^2\right)^{3 / 4}}} \mathrm{Mpc} 
$$

In the data sets, the value $\Omega_{\mathrm{b}, 0} h^2 = 0.0222$ from Planck18\cite{Aghanim:2018eyx} is considered unless otherwise specified.

\subsubsection{Compressed Planck likelihood}

We have utilized the approach suggested by \cite{Arendse_2020} in the compressed Planck likelihood to estimate the baryon physical density $\omega_b = \Omega_b h^2$ and the two shift parameters. For a more detailed explanation, please refer to their Appendix A. The two shift parameters are given by:

$$
\theta_*=r_s\left(z_{\text {dec }}\right) / D_A\left(z_{\text {dec }}\right), \quad \mathcal{R}=\sqrt{\Omega_M H_0^2} D_A\left(z_{\text {dec }}\right),
$$

Here, $z_{d e c}$ represents the redshift at decoupling, and $D_A$ denotes the comoving angular diameter distance. We also confirmed that the compressed likelihood produces the standard Planck constraints for a flat $\Lambda C D M$ model, as stated in \cite{Arendse_2020}.

\subsection{Observational Constraints}\label{sec:bestfit}
 We assumed flat priors for both the cosmological  ($100~\omega_{b}:[1.9,2.5], \omega_{cdm}:[0.095,0.145]$) and model parameter ($\beta:[-1,1]$). The $\alpha$ parameters which are related to the parametrization of the potential were assigned a prior of $[-2,2]$. Based on \cite{Sabti:2021xvh}, we set the sound horizon angular scale $\theta_{s}$ to the Planck CMB value of 1.04110 \cite{Aghanim:2018eyx} and derive the current value of the Hubble parameter $H_0$. Fixing $\theta_{s}$ does not significantly affect the results since it is determined by the acoustic peak angular scales and is mostly independent of the CMB era physics.

\begin{figure*}[h!]
    \centering
    \includegraphics[width=\textwidth]{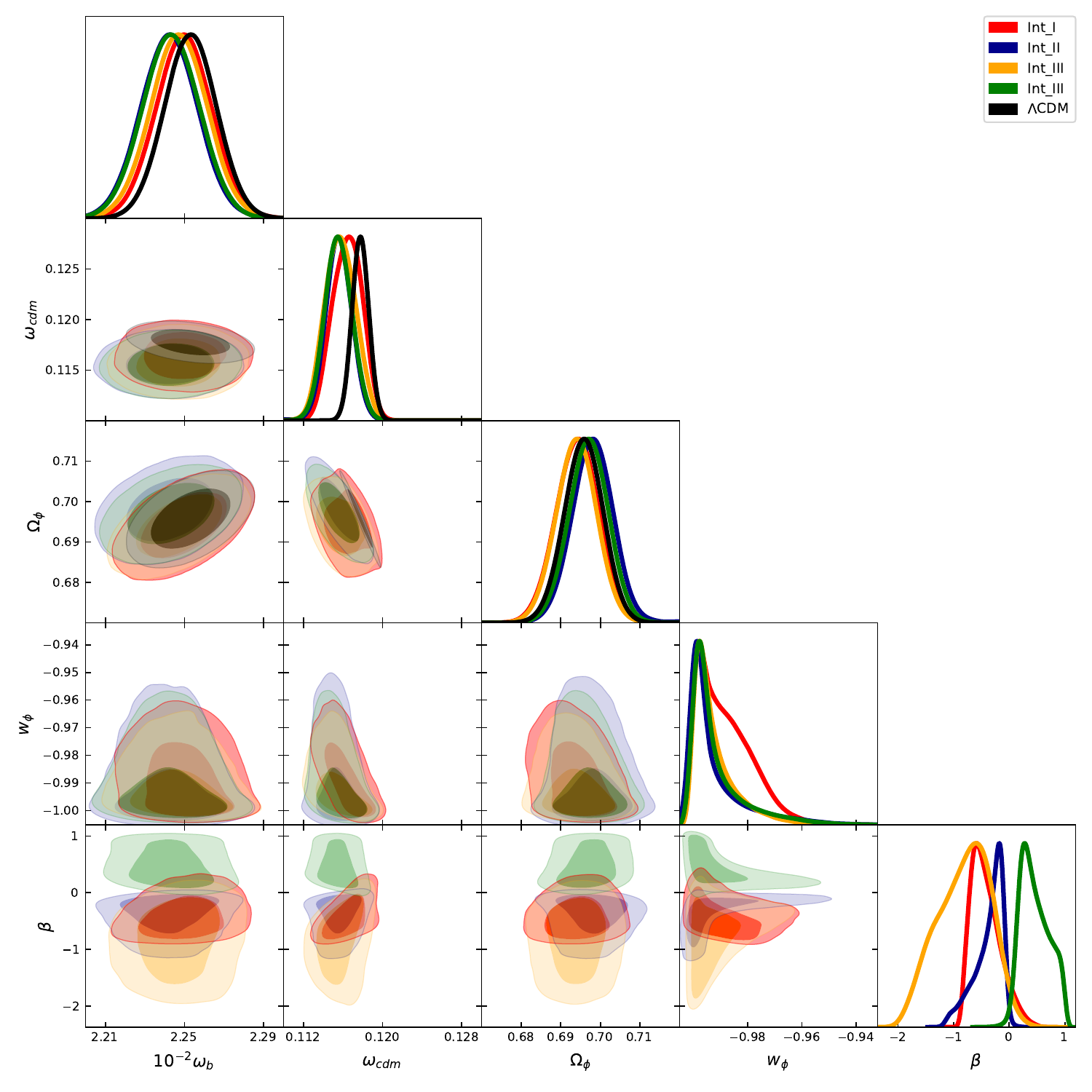}
    \caption{A triangular plot shows the constraints on cosmological parameters for four different interactions, with the $\Lambda$CDM model's constraints shown in black for comparison.  }
    \label{fig:cosmo}
\end{figure*}

The constrain at the $68\%$CL on the cosmological parameters together with the corresponding mean value are given in the Table.\ref{Tab:constraint} for the combined data sets of SN-Ia Pantheon compilation\cite{Scolnic:2017caz}, BAO\cite{Alam_2017,deSainteAgathe:2019voe,Cuceu_2019,Kazin_2014}, with the compressed Planck likelihood\cite{Arendse_2020}. In fig.\ref{fig:cosmo} we have shown the 2D and 1D triangular plots of the cosmological parameters  $H_0,100 w_b, w_{cdm}, \Omega_{\phi}, w_\phi, \beta$. A comparison with the $\Lambda CDM$  model has been shown by plotting it in black. For all the interacting models $100 w_b, w_{cdm}$ are lower than the $\Lambda CDM$  model. 

\begin{figure}
    \centering
    \includegraphics[width=\columnwidth]{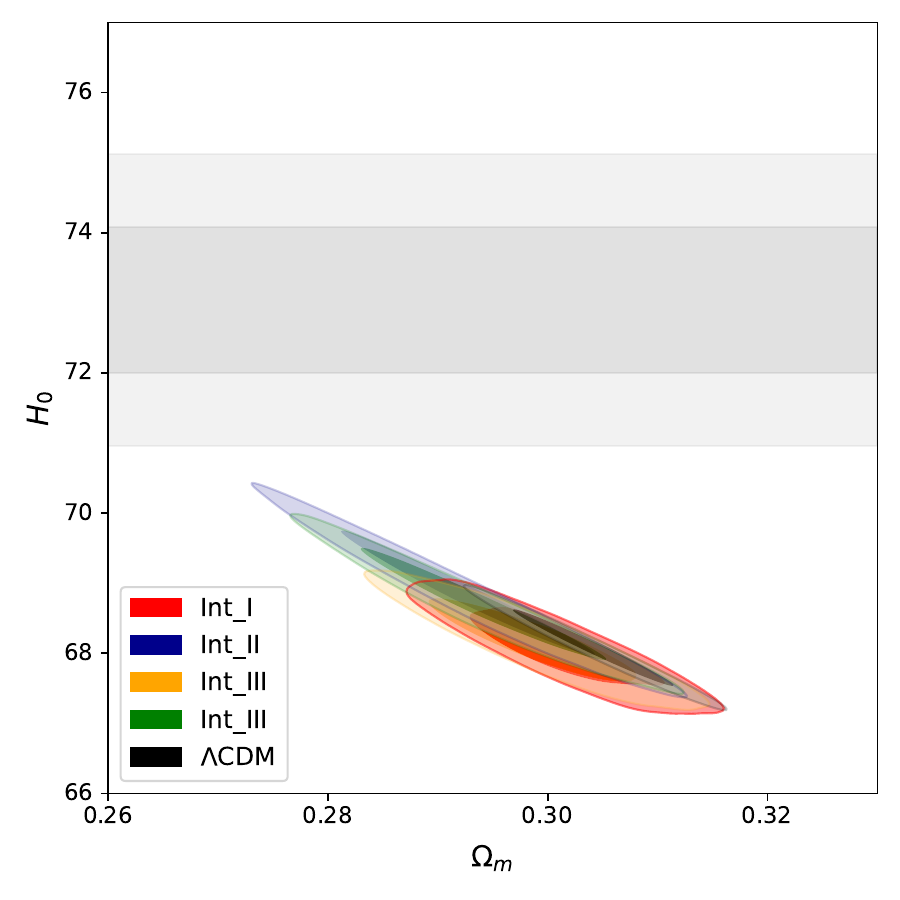}
    \caption{A posterior plot of $H_0$ versus $\Omega_m$ demonstrates how the interacting models perform in resolving the Hubble tension. The horizontal gray band shows the constraint on the $H_0$ from SH0ES measurement\cite{Sh0ES2019}. }
    \label{fig:hubble}
\end{figure}

In Figure~\ref{fig:hubble}, we present a contour plot depicting the posterior distributions of the Hubble parameter $H_0$ and the matter density parameter $\Omega_m$ for all four interacting models together with the $\Lambda$ CDM (in black) for comparison. The horizontal gray regions represent $1\sigma$ and $2\sigma$ constraint on the $H_0$ obtained from the SH0ES collaboration\cite{Sh0ES2019}. Notably, there is some increment in the current value of the Hubble parameter for all four models. To quantify the status of the tension in $H_0$ for these models, we utilize the estimator proposed in \cite{Camarena:2018nbr}, which is given by

\begin{equation}
T_{H0}=\frac{\mid H_0-H_0^{R18}\mid}{\sqrt{\sigma_{H0}^2+\sigma_{\mathrm{loc}}^2}},
\end{equation}

where $T_{H0}$ represents the tension, $H_0$ is the mean of the posterior $p(H_0)$, $\sigma_{H0}^2$ is the variance of the posterior $p(H_0)$, and $\sigma_{\mathrm{loc}}^2$ represents the uncertainty arising from local measurements.  For interaction I, $T_{H0}\simeq 3.83 \sigma$, for interaction II, $T_{H0}\simeq 3.51\sigma$, for interaction III, $T_{H0}\simeq 4.3\sigma$ and interaction IV $T_{H0}\simeq 3.64\sigma$. From the result of this estimator the performance of interaction II is better in terms of solving the Hubble tension.

Also, note from Table.\ref{Tab:constraint} that the value of the coupling parameter $\beta$ is negative for interactions I to III, and positive for interaction IV. This is because interaction IV is linearly dependent on $w_{\phi}$, and since $w_{\phi}$ is presently negative to counterbalance  it, $\beta$ is positive. However, from the signature of the coupling parameter $\beta$ alone, one cannot conclude about the signature of the interaction term $Q$ at present. To determine the signature of $Q$, one also needs to check the signature of $\dot{\phi}$. From equation (\ref{eq:4a}), we can write $\dot{\phi} = \frac{1}{\kappa}\sqrt{6} H \Omega_{\phi}^{1/2} \sin(\frac{1}{2}\cos^{-1}(-w_{\phi}))$, since $w_\phi = -\cos\theta$. It can be easily checked that for all the interacting models, $\dot{\phi} > 0$ once we consider the corresponding mean value of $w_\phi$ from Table \ref{Tab:constraint}, and hence $Q<0$. This indicates a transfer of energy from dark energy to dark matter at present, which is opposite to the expectation for an accelerating universe.

\begin{figure}
    \centering
    \includegraphics[width=\columnwidth]{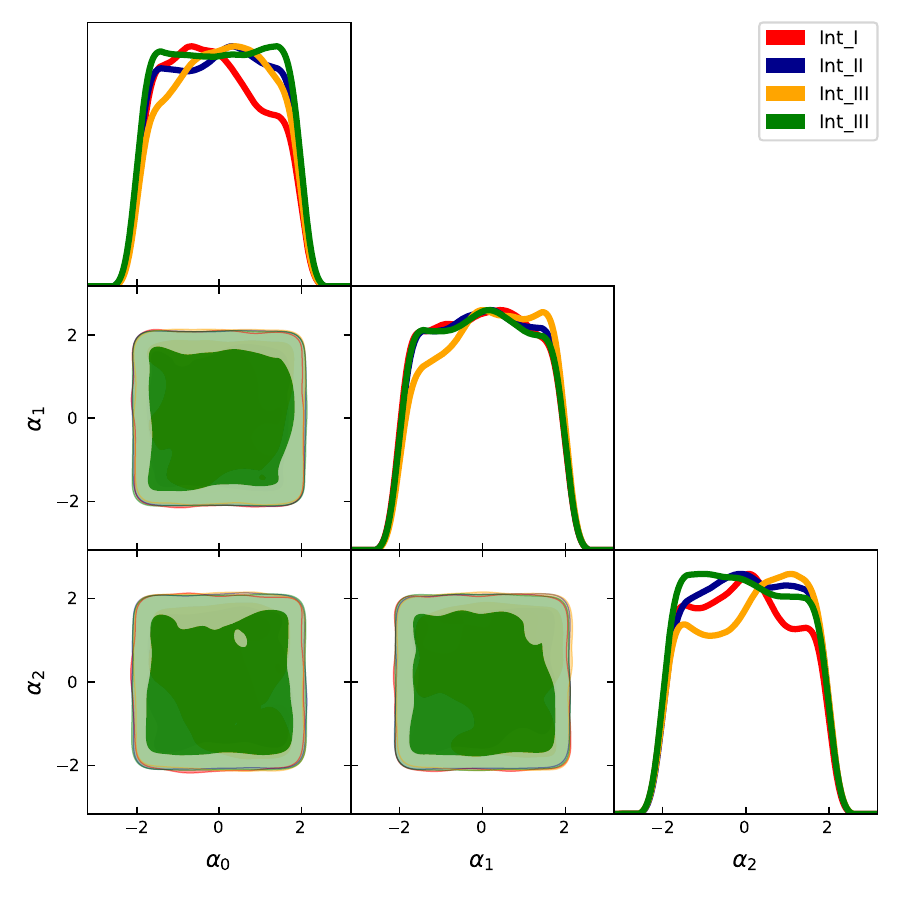}
    \caption{A triangular posterior plot of the $\alpha$ parameters, with a prior range of [-2, 2] for each parameter. One can notice there is no constraint on the $\alpha$ parameters. }
    \label{fig:model}
\end{figure}

Fig. \ref{fig:model} shows the posteriors for the $\alpha$ parameters, and it can be observed that the $\alpha$ parameters remain unconstrained for all the interacting models, which is consistent with previous findings \cite{Roy:2018nce, Roy:2018eug}.

\begin{figure}[h!]
    \centering
    \includegraphics[width=\columnwidth]{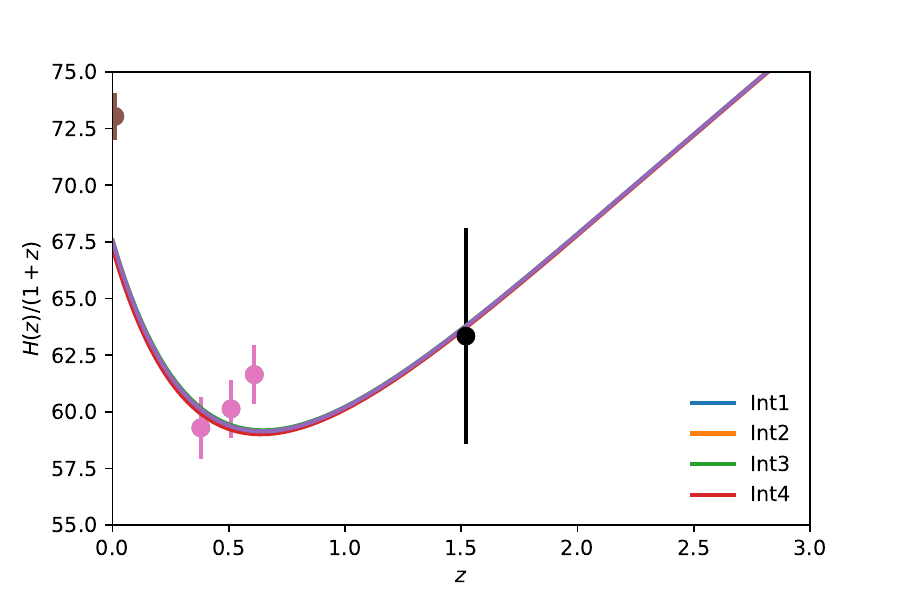}
    \caption{A plot of $H(z)/(1+z)$ for the interacting models, using the same $\alpha$ parameter values as in fig. \ref{fig:eos}, is presented along with observations from the Sh0ES survey \cite{Riess:2019cxk} and Baryon Acoustic Oscillation (BAO) surveys \cite{BOSS:2016wmc,Zarrouk:2018vwy,Blomqvist:2019rah,deSainteAgathe:2019voe} for comparison.}
    \label{fig:H(z)}
\end{figure}

In Fig. \ref{fig:H(z)}, we have plotted the expansion rate of the universe $H(z)/(1+z)$ as a function of $z$. For comparison, we have also shown observational data from Sh0ES \cite{Riess:2019cxk} and BAO observations \cite{BOSS:2016wmc, Zarrouk:2018vwy, Blomqvist:2019rah, deSainteAgathe:2019voe}. From this plot, it can be seen that these interacting models can replicate the $\Lambda$ CDM model well.


\subsection{Comparison with $\Lambda CDM$}
To be certain about the performance of the interacting models we compared the interacting models to the $\Lambda$CDM model using the Bayes factor, which was calculated as $\ln B_{I \Lambda} = \ln \mathcal{Z}{I} - \ln \mathcal{Z}\Lambda$, where $\mathcal{Z}$ represents the Bayesian evidence and the suffixes $I$ and $\Lambda$ represent the interacting models and $\Lambda$CDM models, respectively. To determine the preference for one model over another, we used Jeffrey's scale. A negative preference was assigned if $\lvert \ln B_{I \Lambda}\rvert < 1$, while positive, moderate, and strong preferences were assigned if $\lvert \ln B_{I \Lambda}\rvert > 1$, $\lvert \ln B_{I \Lambda}\rvert > 2.5$, and $\lvert \ln B_{I \Lambda}\rvert > 5.0$, respectively \cite{Trotta:2005ar}. We used the publicly available code \textsc{MCEvidence}~\cite{Heavens:2017afc} to directly calculate the Bayes factor from the MCMC chains generated by \textsc{MontePython}. Our analysis shows positive evidence for all of  the interacting models over $\Lambda$CDM (see Table.\ref{Tab:constraint}). According to Jeffrey's scale, Int1 and Int3 show positive evidence, whereas Int2 and Int4 show moderate evidence over the $\Lambda$CDM model.

\begin{table*}[h!]
\centering
\begin{tabular}{l c c c c c}
Parameter& $\Lambda$ CDM & Int1& Int2 & Int3 & Int4 \\
\hline
{\boldmath$10^{-2}\omega_{b}$} & $2.253\pm 0.014            $ & $2.250\pm 0.017$ & $2.243\pm 0.015$ & $2.248\pm 0.017$ & $2.243\pm 0.015$\\
{\boldmath$\omega_{cdm}$} & $0.11775\pm 0.00094        $&$0.1164^{+0.0016}_{-0.0014}$ & $0.1154^{+0.0014}_{-0.0011}$ & $0.1157\pm 0.0018$ & $0.1155\pm 0.0015$\\
{\boldmath${\beta}$} & - &$-0.42^{+0.15}_{-0.32}$ & $-0.37^{+0.30}_{-0.11}$ & $-0.81^{+0.55}_{-0.41}$ & $0.46^{+0.18}_{-0.33}$\\
{\boldmath$\Omega_{\phi}$} &$0.6958\pm 0.0053          $& $0.6946^{+0.0048}_{-0.0055}$ & $0.6986^{+0.0045}_{-0.0055}$ & $0.6943^{+0.0045}_{-0.0052}$ & $0.6971\pm 0.0055$\\
{\boldmath$\Omega_m$}& $0.3041\pm 0.0053          $ & $0.3009^{+0.0059}_{-0.0053}$ & $0.2917^{+0.0080}_{-0.0063}$ & $0.2985^{+0.0063}_{-0.0057}$ & $0.2939\pm 0.0077$\\
{\boldmath$H_0$}& $68.08\pm 0.41             $ & $68.10^{+0.35}_{-0.41}$ & $68.92^{+0.43}_{-0.65}$ & $68.20^{+0.35}_{-0.42}$ & $68.68\pm 0.59$\\
{\boldmath$w{\phi}$} & -1&$-0.9884^{+0.0052}_{-0.012}$ & $-0.99204^{+0.00038}_{-0.0086}$ & $-0.9922^{+0.0015}_{-0.0082}$ & $-0.99146^{+0.00066}_{-0.0089}$\\
{\boldmath$\lvert \ln B_{I \Lambda} \rvert $}&- & $1.6$ & $2.7$ & $2.4$ & $2.7$ \\
\hline
\end{tabular}
\caption{$68\%$CL constraint on the cosmological parameters from the combining data sets of Pantheon supernova survey \cite{Scolnic:2017caz}, Baryon Acoustic Oscillations from the BOSS DR12 survey \cite{Alam_2017}, Lyman-alpha forest data from the eBOSS DR14 survey \cite{Cuceu_2019}, and the WiggleZ galaxy survey \cite{Kazin_2014}, along with the compressed Planck likelihood. \label{Tab:constraint}}
\end{table*}

\section{Conclusions} \label{conclusion} 

In this study, we have explored the possibility of the interacting quintessence dark energy models as the alternative to the $\Lambda$CDM model. The choice of the type of interaction is phenomenological and also mathematically motivated to close the autonomous system. A general parametrization of the quintessence potential has been considered to study a large class of the potential in a single setup.

The Einstein field equations were reformulated into autonomous equations, and the models were implemented in the Boltzmann code CLASS and evaluated against recent cosmological observations. The state-of-the-art cosmological data sets are used to constrain the cosmological parameters of these models. Our finding suggests that even if there is a shift in the best-fit value of the $H_0$ towards the higher value but these models are far from solving the Hubble tension.

The cosmological evolution of the interacting models can well replicate the $\Lambda$CDM model at the background level but to be certain the results were compared using Bayesian model selection based on the Bayes factor and Jeffreys scale, providing insight into the relative strengths and weaknesses of these models. The study found that the interacting models are more favored by observations compared to $\Lambda$CDM. Out of the four interactions considered here, two show positive evidence, and the other two show moderate evidence over $\Lambda$CDM. We must mention here that current work is limited to the background level, in principle one should consider the evolution of the linear perturbations for a better understanding of these models which will be presented elsewhere in the future.

\section{Acknowledgement}
The author acknowledges the use of the Chalawan High Performance Computing cluster, operated and maintained by the National Astronomical Research Institute of Thailand (NARIT). The research is supported by Mahidol University, Thailand through the research project MU-MRC-MGR 04/2565.

\subsection*{Data Availability Statement:} No Data associated in the manuscript.

\bibliographystyle{plain}
\bibliography{qeos}

\begin{thebibliography}{10}

\bibitem{DES2018}
T.~M.~C. Abbott et~al.
\newblock {Dark Energy Survey year 1 results: Cosmological constraints from
  galaxy clustering and weak lensing}.
\newblock {\em Phys. Rev. D}, 98(4):043526, 2018.

\bibitem{Aghanim:2018eyx}
N.~Aghanim et~al.
\newblock {Planck 2018 results. VI. Cosmological parameters}.
\newblock 2018.

\bibitem{Planck2020}
N.~Aghanim et~al.
\newblock {Planck 2018 results. VI. Cosmological parameters}.
\newblock {\em Astron. Astrophys.}, 641:A6, 2020.
\newblock [Erratum: Astron.Astrophys. 652, C4 (2021)].

\bibitem{ahn2012ninth}
Christopher~P Ahn, Rachael Alexandroff, Carlos~Allende Prieto, Scott~F
  Anderson, Timothy Anderton, Brett~H Andrews, {\'E}ric Aubourg, Stephen
  Bailey, Eduardo Balbinot, Rory Barnes, et~al.
\newblock The ninth data release of the sloan digital sky survey: first
  spectroscopic data from the sdss-iii baryon oscillation spectroscopic survey.
\newblock {\em The Astrophysical Journal Supplement Series}, 203(2):21, 2012.

\bibitem{BAO2017}
Shadab Alam, Metin Ata, Stephen Bailey, Florian Beutler, Dmitry Bizyaev,
  Jonathan~A. Blazek, Adam~S. Bolton, Joel~R. Brownstein, Angela Burden,
  Chia-Hsun Chuang, and et~al.
\newblock The clustering of galaxies in the completed sdss-iii baryon
  oscillation spectroscopic survey: cosmological analysis of the dr12 galaxy
  sample.
\newblock {\em Monthly Notices of the Royal Astronomical Society},
  470(3):2617–2652, Mar 2017.

\bibitem{Alam_2017}
Shadab Alam, Metin Ata, Stephen Bailey, Florian Beutler, Dmitry Bizyaev,
  Jonathan~A. Blazek, Adam~S. Bolton, Joel~R. Brownstein, Angela Burden,
  Chia-Hsun Chuang, and et~al.
\newblock The clustering of galaxies in the completed sdss-iii baryon
  oscillation spectroscopic survey: cosmological analysis of the dr12 galaxy
  sample.
\newblock {\em Monthly Notices of the Royal Astronomical Society},
  470(3):2617–2652, Mar 2017.

\bibitem{BBN2021}
Shadab Alam, Marie Aubert, Santiago Avila, Christophe Balland, Julian~E.
  Bautista, Matthew~A. Bershady, Dmitry Bizyaev, Michael~R. Blanton, Adam~S.
  Bolton, Jo~Bovy, and et~al.
\newblock Completed sdss-iv extended baryon oscillation spectroscopic survey:
  Cosmological implications from two decades of spectroscopic surveys at the
  apache point observatory.
\newblock {\em Physical Review D}, 103(8), Apr 2021.

\bibitem{BOSS:2016wmc}
Shadab Alam et~al.
\newblock {The clustering of galaxies in the completed SDSS-III Baryon
  Oscillation Spectroscopic Survey: cosmological analysis of the DR12 galaxy
  sample}.
\newblock {\em Mon. Not. Roy. Astron. Soc.}, 470(3):2617--2652, 2017.

\bibitem{amendola2000coupled}
Luca Amendola.
\newblock Coupled quintessence.
\newblock {\em Physical Review D}, 62(4):043511, 2000.

\bibitem{Amendola:1999er}
Luca Amendola.
\newblock {Coupled quintessence}.
\newblock {\em Phys. Rev. D}, 62:043511, 2000.

\bibitem{amendola2010dark}
Luca Amendola and Shinji Tsujikawa.
\newblock {\em Dark energy: theory and observations}.
\newblock Cambridge University Press, 2010.

\bibitem{Arendse_2020}
Nikki Arendse, Radosław~J. Wojtak, Adriano Agnello, Geoff C.-F. Chen,
  Christopher~D. Fassnacht, Dominique Sluse, Stefan Hilbert, Martin Millon,
  Vivien Bonvin, Kenneth~C. Wong, and et~al.
\newblock Cosmic dissonance: are new physics or systematics behind a short
  sound horizon?
\newblock {\em Astronomy \& Astrophysics}, 639:A57, Jul 2020.

\bibitem{Armendariz2001}
C.~Armendariz-Picon, V.~Mukhanov, and P.~J. Steinhardt.
\newblock k-essence as a model for dark energy.
\newblock {\em Physical Review Letters}, 85(15):4438--4441, 2001.

\bibitem{Planck:2014loa}
M.~Arnaud et~al.
\newblock {Planck intermediate results. XXXI. Microwave survey of Galactic
  supernova remnants}.
\newblock {\em Astron. Astrophys.}, 586:A134, 2016.

\bibitem{Bahamonde:2017ize}
Sebastian Bahamonde, Christian~G. Böhmer, Sante Carloni, Edmund~J. Copeland,
  Wei Fang, and Nicola Tamanini.
\newblock {Dynamical systems applied to cosmology: dark energy and modified
  gravity}.
\newblock {\em Phys. Rept.}, 775-777:1--122, 2018.

\bibitem{Bamba:2012cp}
Kazuharu Bamba, Salvatore Capozziello, Shin'ichi Nojiri, and Sergei~D.
  Odintsov.
\newblock {Dark energy cosmology: the equivalent description via different
  theoretical models and cosmography tests}.
\newblock {\em Astrophys. Space Sci.}, 342:155--228, 2012.

\bibitem{Banerjee:2020xcn}
Aritra Banerjee, Haiying Cai, Lavinia Heisenberg, Eoin~\'O. Colg\'ain, M.~M.
  Sheikh-Jabbari, and Tao Yang.
\newblock {Hubble sinks in the low-redshift swampland}.
\newblock {\em Phys. Rev. D}, 103(8):L081305, 2021.

\bibitem{BAO2011}
Florian Beutler, Chris Blake, Matthew Colless, D.~Heath Jones, Lister
  Staveley-Smith, Lachlan Campbell, Quentin Parker, Will Saunders, and Fred
  Watson.
\newblock The 6df galaxy survey: baryon acoustic oscillations and the local
  hubble constant.
\newblock {\em Monthly Notices of the Royal Astronomical Society},
  416(4):3017–3032, Jul 2011.

\bibitem{Blas:2011rf}
Diego Blas, Julien Lesgourgues, and Thomas Tram.
\newblock {The Cosmic Linear Anisotropy Solving System (CLASS) II:
  Approximation schemes}.
\newblock {\em JCAP}, 1107:034, 2011.

\bibitem{Blomqvist:2019rah}
Michael Blomqvist et~al.
\newblock {Baryon acoustic oscillations from the cross-correlation of
  Ly$\alpha$ absorption and quasars in eBOSS DR14}.
\newblock {\em Astron. Astrophys.}, 629:A86, 2019.

\bibitem{Boehmer:2008av}
Christian~G. Boehmer, Gabriela Caldera-Cabral, Ruth Lazkoz, and Roy Maartens.
\newblock {Dynamics of dark energy with a coupling to dark matter}.
\newblock {\em Phys. Rev. D}, 78:023505, 2008.

\bibitem{Brinckmann:2018cvx}
Thejs Brinckmann and Julien Lesgourgues.
\newblock {MontePython 3: boosted MCMC sampler and other features}.
\newblock 2018.

\bibitem{Cai:2004dk}
Rong-Gen Cai and Anzhong Wang.
\newblock {Cosmology with interaction between phantom dark energy and dark
  matter and the coincidence problem}.
\newblock {\em JCAP}, 03:002, 2005.

\bibitem{Caldera-Cabral:2008yyo}
Gabriela Caldera-Cabral, Roy Maartens, and L.~Arturo Urena-Lopez.
\newblock {Dynamics of interacting dark energy}.
\newblock {\em Phys. Rev. D}, 79:063518, 2009.

\bibitem{Camarena:2018nbr}
David Camarena and Valerio Marra.
\newblock {Impact of the cosmic variance on $H_0$ on cosmological analyses}.
\newblock {\em Phys. Rev. D}, 98(2):023537, 2018.

\bibitem{chimento2010linear}
Luis~P Chimento.
\newblock Linear and nonlinear interactions in the dark sector.
\newblock {\em Physical Review D}, 81(4):043525, 2010.

\bibitem{copeland2006dynamics}
Edmund~J Copeland, Mohammad Sami, and Shinji Tsujikawa.
\newblock Dynamics of dark energy.
\newblock {\em International Journal of Modern Physics D}, 15(11):1753--1935,
  2006.

\bibitem{Costa2017}
A.~Costa and P.~G. Ferreira.
\newblock Hubble tension and interacting dark energy.
\newblock {\em Journal of Cosmology and Astroparticle Physics}, 2017(12):013,
  2017.

\bibitem{Cuceu_2019}
Andrei Cuceu, James Farr, Pablo Lemos, and Andreu Font-Ribera.
\newblock Baryon acoustic oscillations and the hubble constant: past, present
  and future.
\newblock {\em Journal of Cosmology and Astroparticle Physics},
  2019(10):044–044, Oct 2019.

\bibitem{deSainteAgathe:2019voe}
Victoria de~Sainte~Agathe et~al.
\newblock {Baryon acoustic oscillations at z = 2.34 from the correlations of
  Ly$\alpha$ absorption in eBOSS DR14}.
\newblock {\em Astron. Astrophys.}, 629:A85, 2019.

\bibitem{DiValentino2019}
E.~Di~Valentino, A.~Melchiorri, and J.~Silk.
\newblock Cosmological constraints from the combination of latest data sets:
  the role of dark energy interactions.
\newblock {\em The European Physical Journal C}, 79(2):139, 2019.

\bibitem{DiValentino:2019ffd}
Eleonora Di~Valentino, Alessandro Melchiorri, Olga Mena, and Sunny Vagnozzi.
\newblock {Interacting dark energy in the early 2020s: A promising solution to
  the $H_0$ and cosmic shear tensions}.
\newblock {\em Phys. Dark Univ.}, 30:100666, 2020.

\bibitem{DiValentino:2019jae}
Eleonora Di~Valentino, Alessandro Melchiorri, Olga Mena, and Sunny Vagnozzi.
\newblock {Nonminimal dark sector physics and cosmological tensions}.
\newblock {\em Phys. Rev. D}, 101(6):063502, 2020.

\bibitem{farrar2004interacting}
Glennys~R Farrar and P~James~E Peebles.
\newblock Interacting dark matter and dark energy.
\newblock {\em The Astrophysical Journal}, 604(1):1, 2004.

\bibitem{Heavens:2017afc}
Alan Heavens, Yabebal Fantaye, Arrykrishna Mootoovaloo, Hans Eggers, Zafiirah
  Hosenie, Steve Kroon, and Elena Sellentin.
\newblock {Marginal Likelihoods from Monte Carlo Markov Chains}.
\newblock 4 2017.

\bibitem{Hussain:2022dhp}
Saddam Hussain, Saikat Chakraborty, Nandan Roy, and Kaushik Bhattacharya.
\newblock {Dynamical systems analysis of tachyon-dark-energy models from a new
  perspective}.
\newblock {\em Phys. Rev. D}, 107(6):063515, 2023.

\bibitem{Jesus:2020tby}
J.~F. Jesus, A.~A. Escobal, D.~Benndorf, and S.~H. Pereira.
\newblock {Can dark matter\textendash{}dark energy interaction alleviate the
  cosmic coincidence problem?}
\newblock {\em Eur. Phys. J. C}, 82(3):273, 2022.

\bibitem{Kazin_2014}
Eyal~A. Kazin, Jun Koda, Chris Blake, Nikhil Padmanabhan, Sarah Brough, Matthew
  Colless, Carlos Contreras, Warrick Couch, Scott Croom, Darren~J. Croton, and
  et~al.
\newblock The wigglez dark energy survey: improved distance measurements to z =
  1 with reconstruction of the baryonic acoustic feature.
\newblock {\em Monthly Notices of the Royal Astronomical Society},
  441(4):3524–3542, May 2014.

\bibitem{Khyllep:2021wjd}
Wompherdeiki Khyllep, Jibitesh Dutta, Spyros Basilakos, and Emmanuel~N.
  Saridakis.
\newblock {Background evolution and growth of structures in interacting dark
  energy scenarios through dynamical system analysis}.
\newblock {\em Phys. Rev. D}, 105(4):043511, 2022.

\bibitem{krause2017dark}
E.~Krause et~al.
\newblock {Dark Energy Survey Year 1 Results: Multi-Probe Methodology and
  Simulated Likelihood Analyses}.
\newblock 6 2017.

\bibitem{Krishnan:2020vaf}
C.~Krishnan, E.~\'O. Colg\'ain, M.~M. Sheikh-Jabbari, and Tao Yang.
\newblock {Running Hubble Tension and a H0 Diagnostic}.
\newblock {\em Phys. Rev. D}, 103(10):103509, 2021.

\bibitem{Kumar2020}
S.~Kumar, S.~Kumar, K.~Liao, and Y.~Wang.
\newblock Interacting dark energy models with a logarithmic interaction term
  and their implications on the hubble tension.
\newblock {\em Astrophysics and Space Science}, 365(6):207, 2020.

\bibitem{Kumar:2019wfs}
Suresh Kumar, Rafael~C. Nunes, and Santosh~Kumar Yadav.
\newblock {Dark sector interaction: a remedy of the tensions between CMB and
  LSS data}.
\newblock {\em Eur. Phys. J. C}, 79(7):576, 2019.

\bibitem{Lee:2022cyh}
Bum-Hoon Lee, Wonwoo Lee, Eoin~\'O. Colg\'ain, M.~M. Sheikh-Jabbari, and
  Somyadip Thakur.
\newblock {Is local H $_{0}$ at odds with dark energy EFT?}
\newblock {\em JCAP}, 04(04):004, 2022.

\bibitem{Lesgourgues:2011rg}
Julien Lesgourgues.
\newblock {The Cosmic Linear Anisotropy Solving System (CLASS) III: Comparision
  with CAMB for LambdaCDM}.
\newblock 2011.

\bibitem{Lesgourgues:2011rh}
Julien Lesgourgues and Thomas Tram.
\newblock {The Cosmic Linear Anisotropy Solving System (CLASS) IV: efficient
  implementation of non-cold relics}.
\newblock {\em JCAP}, 1109:032, 2011.

\bibitem{DES:2018rjw}
E.~Macaulay et~al.
\newblock {First Cosmological Results using Type Ia Supernovae from the Dark
  Energy Survey: Measurement of the Hubble Constant}.
\newblock {\em Mon. Not. Roy. Astron. Soc.}, 486(2):2184--2196, 2019.

\bibitem{mangano2003coupled}
G~Mangano, Gennaro Miele, and V~Pettorino.
\newblock Coupled quintessence and the coincidence problem.
\newblock {\em Modern Physics Letters A}, 18(12):831--842, 2003.

\bibitem{EUCLID:2020syl}
M.~Martinelli et~al.
\newblock {Euclid: Forecast constraints on the cosmic distance duality relation
  with complementary external probes}.
\newblock {\em Astron. Astrophys.}, 644:A80, 2020.

\bibitem{Meszaros:2002np}
Attila Meszaros.
\newblock {On the Reality of the accelerating universe}.
\newblock {\em Astrophys. J.}, 580:12--15, 2002.

\bibitem{padmanabhan2006dark}
T~Padmanabhan.
\newblock Dark energy: mystery of the millennium.
\newblock In {\em AIP Conference Proceedings}, volume 861, pages 179--196.
  American Institute of Physics, 2006.

\bibitem{pan2015analytic}
Supriya Pan, Subhra Bhattacharya, and Subenoy Chakraborty.
\newblock An analytic model for interacting dark energy and its observational
  constraints.
\newblock {\em Monthly Notices of the Royal Astronomical Society},
  452(3):3038--3046, 2015.

\bibitem{Peebles2003}
P.~J.~E. Peebles and B.~Ratra.
\newblock Quintessence: A review.
\newblock {\em Reviews of Modern Physics}, 75(2):559--606, 2003.

\bibitem{SupernovaCosmologyProject:1998vns}
S.~Perlmutter et~al.
\newblock {Measurements of $\Omega$ and $\Lambda$ from 42 high redshift
  supernovae}.
\newblock {\em Astrophys. J.}, 517:565--586, 1999.

\bibitem{pettorino2008coupled}
Valeria Pettorino and Carlo Baccigalupi.
\newblock Coupled and extended quintessence: theoretical differences and
  structure formation.
\newblock {\em Physical Review D}, 77(10):103003, 2008.

\bibitem{pettorino2005extended}
Valeria Pettorino, Carlo Baccigalupi, and Gianpiero Mangano.
\newblock Extended quintessence with an exponential coupling.
\newblock {\em Journal of Cosmology and Astroparticle Physics}, 2005(01):014,
  2005.

\bibitem{reiss1998supernova}
AG~Reiss et~al.
\newblock Supernova serach team.
\newblock {\em Astron. J}, 116:1009, 1998.

\bibitem{Riess_2022}
Adam~G. Riess, Louise Breuval, Wenlong Yuan, Stefano Casertano, Lucas~M. Macri,
  J.~Bradley Bowers, Dan Scolnic, Tristan Cantat-Gaudin, Richard~I. Anderson,
  and Mauricio~Cruz Reyes.
\newblock Cluster cepheids with high precision gaia parallaxes, low zero-point
  uncertainties, and hubble space telescope photometry.
\newblock {\em The Astrophysical Journal}, 938(1):36, oct 2022.

\bibitem{Sh0ES2019}
Adam~G. Riess, Stefano Casertano, Wenlong Yuan, Lucas~M. Macri, and Dan
  Scolnic.
\newblock Large magellanic cloud cepheid standards provide a 1\% foundation for
  the determination of the hubble constant and stronger evidence for physics
  beyond $\lambda$cdm.
\newblock {\em The Astrophysical Journal}, 876(1):85, May 2019.

\bibitem{Riess:2019cxk}
Adam~G. Riess, Stefano Casertano, Wenlong Yuan, Lucas~M. Macri, and Dan
  Scolnic.
\newblock {Large Magellanic Cloud Cepheid Standards Provide a 1\% Foundation
  for the Determination of the Hubble Constant and Stronger Evidence for
  Physics beyond $\Lambda$CDM}.
\newblock {\em Astrophys. J.}, 876(1):85, 2019.

\bibitem{SupernovaSearchTeam:1998fmf}
Adam~G. Riess et~al.
\newblock {Observational evidence from supernovae for an accelerating universe
  and a cosmological constant}.
\newblock {\em Astron. J.}, 116:1009--1038, 1998.

\bibitem{Roy:2018eug}
Nandan Roy and Kazuharu Bamba.
\newblock {Arbitrariness of potentials in interacting quintessence models}.
\newblock {\em Phys. Rev. D}, 99(12):123520, 2019.

\bibitem{Roy:2018nce}
Nandan Roy, Alma~X. Gonzalez-Morales, and L.~Arturo Urena-Lopez.
\newblock {New general parametrization of quintessence fields and its
  observational constraints}.
\newblock {\em Phys. Rev.}, D98(6):063530, 2018.

\bibitem{roy2022quintessence}
Nandan Roy, Sangita Goswami, and Sudipta Das.
\newblock Quintessence or phantom: study of scalar field dark energy models
  through a general parametrization of the hubble parameter.
\newblock {\em Physics of the Dark Universe}, 36:101037, 2022.

\bibitem{Sabti:2021xvh}
Nashwan Sabti, Julian~B. Mu\~noz, and Diego Blas.
\newblock {Galaxy luminosity function pipeline for cosmology and astrophysics}.
\newblock {\em Phys. Rev. D}, 105(4):043518, 2022.

\bibitem{Sadjadi:2006qp}
H.~Mohseni Sadjadi and M.~Alimohammadi.
\newblock {Cosmological coincidence problem in interactive dark energy models}.
\newblock {\em Phys. Rev. D}, 74:103007, 2006.

\bibitem{Salvatelli2014}
V.~Salvatelli, A.~Marchini, L.~Pogosian, N.~Vittorio, Y.-C. Wu, and J.~Zavala.
\newblock Indications of a late-time interaction in the dark sector.
\newblock {\em Physical Review Letters}, 113(18):181301, 2014.

\bibitem{Scolnic:2017caz}
D.~M. Scolnic et~al.
\newblock {The Complete Light-curve Sample of Spectroscopically Confirmed SNe
  Ia from Pan-STARRS1 and Cosmological Constraints from the Combined Pantheon
  Sample}.
\newblock {\em Astrophys. J.}, 859(2):101, 2018.

\bibitem{tamanini2015phenomenological}
Nicola Tamanini.
\newblock Phenomenological models of dark energy interacting with dark matter.
\newblock {\em Physical Review D}, 92(4):043524, 2015.

\bibitem{Trotta:2005ar}
Roberto Trotta.
\newblock {Applications of Bayesian model selection to cosmological
  parameters}.
\newblock {\em Mon. Not. Roy. Astron. Soc.}, 378:72--82, 2007.

\bibitem{Urena-Lopez:2015gur}
L.~Arturo Ureña-López and Alma~X. Gonzalez-Morales.
\newblock {Towards accurate cosmological predictions for rapidly oscillating
  scalar fields as dark matter}.
\newblock {\em JCAP}, 1607(07):048, 2016.

\bibitem{Urena-Lopez:2020npg}
L.~Arturo Ureña-López and Nandan Roy.
\newblock Generalized tracker quintessence models for dark energy.
\newblock {\em Physical Review D}, 102(6), Sep 2020.

\bibitem{Wang:2016lxa}
B.~Wang, E.~Abdalla, F.~Atrio-Barandela, and D.~Pavon.
\newblock {Dark Matter and Dark Energy Interactions: Theoretical Challenges,
  Cosmological Implications and Observational Signatures}.
\newblock {\em Rept. Prog. Phys.}, 79(9):096901, 2016.

\bibitem{Wang:2018duq}
Deng Wang.
\newblock {The multi-feature universe: Large parameter space cosmology and the
  swampland}.
\newblock {\em Phys. Dark Univ.}, 28:100545, 2020.

\bibitem{Wetterich:1994bg}
Christof Wetterich.
\newblock {The Cosmon model for an asymptotically vanishing time dependent
  cosmological 'constant'}.
\newblock {\em Astron. Astrophys.}, 301:321--328, 1995.

\bibitem{Wong:2019kwg}
Kenneth~C. Wong et~al.
\newblock {H0LiCOW \textendash{} XIII. A 2.4 percent measurement of H0 from
  lensed quasars: 5.3\ensuremath{\sigma} tension between early- and
  late-Universe probes}.
\newblock {\em Mon. Not. Roy. Astron. Soc.}, 498(1):1420--1439, 2020.

\bibitem{Yang:2018euj}
Weiqiang Yang, Supriya Pan, Eleonora Di~Valentino, Rafael~C. Nunes, Sunny
  Vagnozzi, and David~F. Mota.
\newblock {Tale of stable interacting dark energy, observational signatures,
  and the $H_0$ tension}.
\newblock {\em JCAP}, 1809:019, 2018.

\bibitem{Zarrouk:2018vwy}
Pauline Zarrouk et~al.
\newblock {The clustering of the SDSS-IV extended Baryon Oscillation
  Spectroscopic Survey DR14 quasar sample: measurement of the growth rate of
  structure from the anisotropic correlation function between redshift 0.8 and
  2.2}.
\newblock {\em Mon. Not. Roy. Astron. Soc.}, 477(2):1639--1663, 2018.

\bibitem{Zonunmawia:2017ofc}
Hmar Zonunmawia, Wompherdeiki Khyllep, Nandan Roy, Jibitesh Dutta, and Nicola
  Tamanini.
\newblock {Extended Phase Space Analysis of Interacting Dark Energy Models in
  Loop Quantum Cosmology}.
\newblock {\em Phys. Rev. D}, 96(8):083527, 2017.

\end{thebibliography}
\end{document}